\documentclass{svproc}

\usepackage{url}

\usepackage[T1]{fontenc}
\usepackage[utf8]{inputenc}
\usepackage{microtype}

\usepackage{graphicx}
\usepackage{amsmath}
\usepackage{amssymb}
\usepackage{booktabs}
\usepackage{multirow}
\usepackage{array}
\usepackage{enumitem}
\usepackage[dvipsnames]{xcolor}
\usepackage{colortbl}
\usepackage[colorlinks=true,linkcolor=blue,citecolor=blue,urlcolor=blue]{hyperref}
\usepackage{placeins}
\usepackage{float}

\begin{document}
\sloppy

\mainmatter

\title{LLMs Remember First, Forget Last: Dual-Process Interference in Large Language Models}
\titlerunning{LLMs Remember First, Forget Last}

\author{Sourav Chattaraj\inst{1} \and Kanak Raj\inst{1}}
\authorrunning{S. Chattaraj and K. Raj}

\institute{Thomson Reuters, Bengaluru, India\\
\email{\{sourav.chattaraj, kanak.raj\}@thomsonreuters.com}}

\maketitle

\begin{abstract}
Large language models can process millions of tokens, yet how they handle conflicting information within context remains poorly understood. From patient health logs tracking evolving vital signs to legal documents with superseding clauses, real-world applications routinely require models to retrieve specific values from streams of semantically similar, competing updates. We adapt classical interference paradigms from cognitive psychology to compare retroactive interference (RI; recalling initial values after updates) and proactive interference (PI; recalling recent values despite competing prior encodings) across 39 LLMs spanning 1B to 2.5T parameters. Every model exhibits the same pattern: PI causes substantially greater degradation than RI ($d = 1.73$), the opposite of the typical human finding where new information more readily disrupts old. Four lines of evidence indicate that RI and PI engage distinct mechanisms: model size predicts RI resistance but not PI; the two scores correlate only weakly; error analysis reveals qualitatively different failure profiles, with RI failures reflecting passive retrieval failure and PI failures reflecting active primacy intrusion; and hallucination rates remain below 1\% in both conditions. An adaptive bootstrap design further shows that RI decays smoothly while PI collapses bimodally at high interference, consistent with gradual dilution versus winner-take-all attention dynamics. These findings characterize a systematic primacy bias in transformer attention, with implications for reliable deployment in settings where in-context information evolves over time.
\end{abstract}

\section{Introduction}

\begin{figure}[t]
\centering
\includegraphics[width=\textwidth]{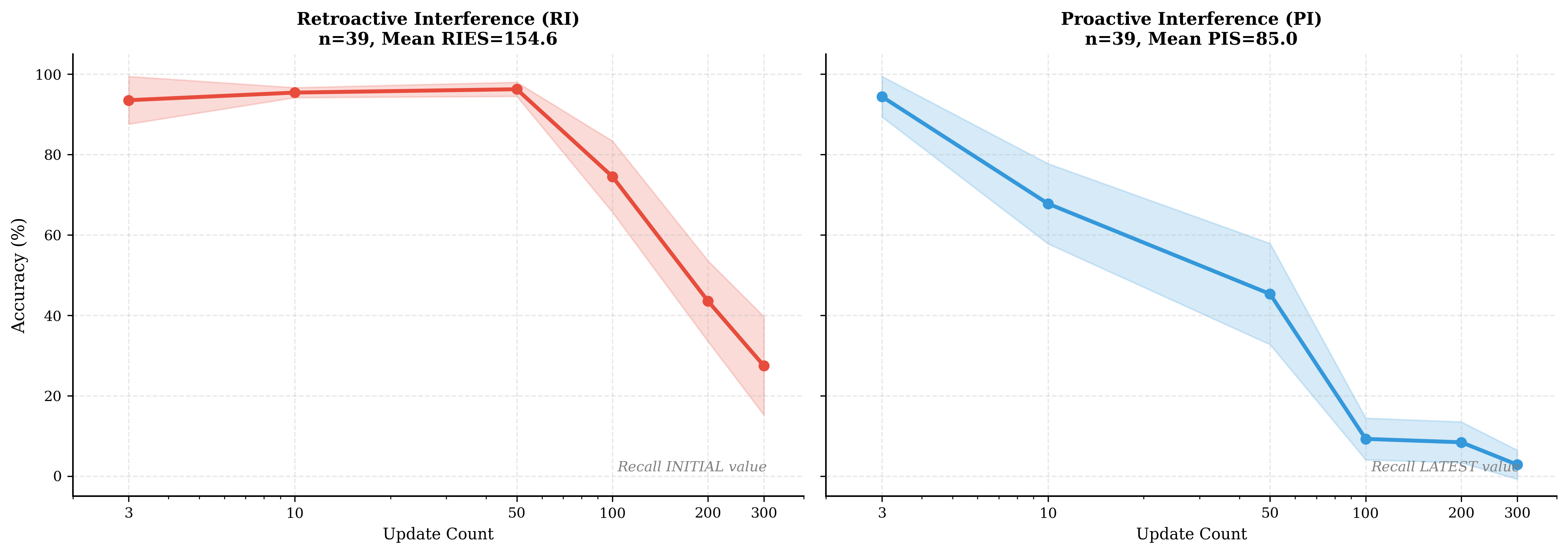}
\caption{\textbf{Interference Asymmetry Across 39 LLMs.} RI accuracy (left) stays near ceiling through $N$=50 before declining to 25\% at $N$=300; PI accuracy (right) drops from 94\% to 3\%, near-total failure. Every model shows PI $>$ RI, the inverse of the dominant human pattern \cite{macleod2024interference}. Shaded bands indicate 95\% bootstrap CIs.}
\label{fig:ri_vs_pi}
\end{figure}

In human cognition, newly acquired information tends to displace older memories, a phenomenon known as \textit{retroactive interference} (RI). Its complement, \textit{proactive interference} (PI), occurs when previously learned material impedes the acquisition or retrieval of new information. Over a century of experimental work indicates that RI generally exerts a stronger effect than PI in human paired-associate learning: recent knowledge overwrites old more than old blocks new \cite{jenkins1924obliviscence,underwood1957interference}.

Consider a medical AI tracking blood pressure updates throughout a 12-hour emergency visit. Asking ``What was the initial reading?'' (RI) and ``What is the current reading?'' (PI) demand opposite memory operations, and our findings suggest LLMs handle them very differently.

Despite advances in few-shot learning \cite{brown2020language,achiam2023gpt4} and context windows expanding to millions of tokens, how LLMs handle memory interference remains poorly understood. We adapted classical interference paradigms to test 39 LLMs. \textbf{Every model shows the opposite pattern}: PI dominates RI universally across all architectures, scales, and training regimes (Figure~\ref{fig:ri_vs_pi}).

This inversion points to distinct computational structure. Under a unified ``memory capacity'' account, RI and PI performance should correlate, yet they do not. Scale predicts RI resistance but leaves PI unaffected, and the two failure modes are qualitatively different: RI errors are passive omissions, while PI errors are active intrusions from earlier values. In neither case do models fabricate; they confuse positions within the material they have seen.

These dissociations echo the consolidation--retrieval distinction long recognized in cognitive science \cite{wixted2004psychology}. RI appears to probe whether initial encodings survive subsequent overwriting, a capacity-limited process that improves with scale. PI probes whether attention can prioritize recent information over entrenched earlier encodings, an architectural property that does not improve simply by adding parameters.

Prior work on LLM memory has addressed only PI \cite{wang2025unable}, leaving the RI side of the asymmetry uncharted. Yet many real-world tasks, from tracking the original terms of a contract to recalling a patient's admission diagnosis, depend on exactly this capability. Clarifying how RI and PI diverge is therefore important for principled model selection and deployment.

\paragraph{Contributions.}
\begin{enumerate}[itemsep=2pt,topsep=3pt,parsep=0pt]
    \item \textbf{Evidence consistent with dual-process memory}: RI and PI scores are only weakly associated ($r = 0.21$) and scale in opposite directions with model size, pointing to a consolidation-driven process behind RI and a retrieval-driven process behind PI.
    \item \textbf{Inverted interference profile}: Every tested LLM exhibits PI $>$ RI ($d = 1.73$), the reverse of the human pattern in which RI typically dominates, providing evidence that transformer self-attention confers a primacy advantage with no obvious parallel in biological memory.
    \item \textbf{Cognitive failure taxonomy}: The two interference types produce qualitatively different errors (passive retrieval failures for RI, active primacy intrusions for PI), with hallucination below 1\% in both cases, indicating positional confusion rather than fabrication.
    \item \textbf{Statistical stability asymmetry}: Under an adaptive bootstrap design, RI accuracy degrades gradually across sessions, whereas PI accuracy at high interference alternates bimodally between near-ceiling and near-floor, consistent with smooth representational dilution for RI and winner-take-all attention dynamics for PI.
\end{enumerate}

\section{Background}

\paragraph{Interference in Human Memory.}
Since the pioneering experiments of Jenkins and Dallenbach~\cite{jenkins1924obliviscence} and the systematic reviews of Underwood~\cite{underwood1957interference}, RI and PI have been recognized as core constraints on human memory. Serial-recall studies reveal both primacy and recency gradients, with recency typically dominating immediate tests \cite{murdock1962serial}. Which interference type proves stronger depends on the specifics of the paradigm \cite{macleod2024interference}: RI tends to predominate in immediate paired-associate tasks, though extended prior learning or categorical shifts can tip the balance toward PI \cite{wickens1970encoding}. Individual differences in working memory capacity further modulate susceptibility to interference \cite{engle2002working,kane2001working}. Contemporary accounts draw a key theoretical line between two sources of forgetting \cite{wixted2004psychology,anderson1994remembering}: \textit{consolidation-based} interference, in which new encodings overwrite memories that have not yet stabilized (RI), and \textit{retrieval-based} interference, in which accumulated traces compete at the point of recall (PI). We adopt this consolidation--retrieval framework as the lens through which to interpret interference patterns in LLMs.

\paragraph{Memory Interference in LLMs.}
It is important to distinguish \textit{in-context interference} (competition among items presented within a single forward pass) from \textit{catastrophic forgetting}, the well-studied phenomenon in which gradient updates during training erode previously acquired knowledge \cite{mccloskey1989catastrophic,french1999catastrophic}. Our focus is on the former: whether models can selectively access specific values when semantically related alternatives compete for retrieval.

Most existing long-context benchmarks test a model's ability to locate a single target embedded in a large passage \cite{kamradt2023needle,bai2024longbench,modarressi2025nolima,fu2025absencebench,ling2025longreason}. Because these tasks mix search difficulty with the volume of surrounding material, they do not isolate what happens when multiple semantically related values compete for the same retrieval cue. Separately, known positional biases in transformer attention, including degraded recall for middle positions \cite{liu2024lost} and disproportionate weight on early tokens \cite{xiao2024efficient}, have not been examined in the context of interference.

Wang and Sun~\cite{wang2025unable} took an important step by introducing a controlled PI paradigm for LLMs, demonstrating that parameter count predicts PI resistance ($R^2 = 0.26$) whereas context window size does not. The complementary question, whether models can still retrieve initial facts after encountering competing updates (RI), has remained open. Although scaling laws predict broad capability gains with model size \cite{kaplan2020scaling,hoffmann2022chinchilla}, it is unknown whether RI and PI benefit equally. The answer matters: if the two forms of interference rely on distinct mechanisms (consolidation vs.\ retrieval), models may be selectively vulnerable to one or the other; if they share a single underlying capacity, the two scores should co-vary. We address this question by administering both RI and PI probes on identical stimulus sequences to 39 LLMs using the AB-AC paradigm \cite{barnes1959fate,underwood1957interference}.

\section{Methods}

\paragraph{Task Design.}
We adapt the classic AB-AC paired-associate interference paradigm \cite{barnes1959fate,underwood1957interference} to probe LLM memory. Each category appears $N$ times in total ($N \in \{3, 10, 50, 100, 200, 300\}$): once as an initial assignment (e.g., \textit{visual art: impressionism}) and $N{-}1$ times as interleaved updates, each rebinding the category to a new value (e.g., \textit{visual art: baroque}).

\begin{figure}[t]
\centering
\includegraphics[width=0.85\textwidth]{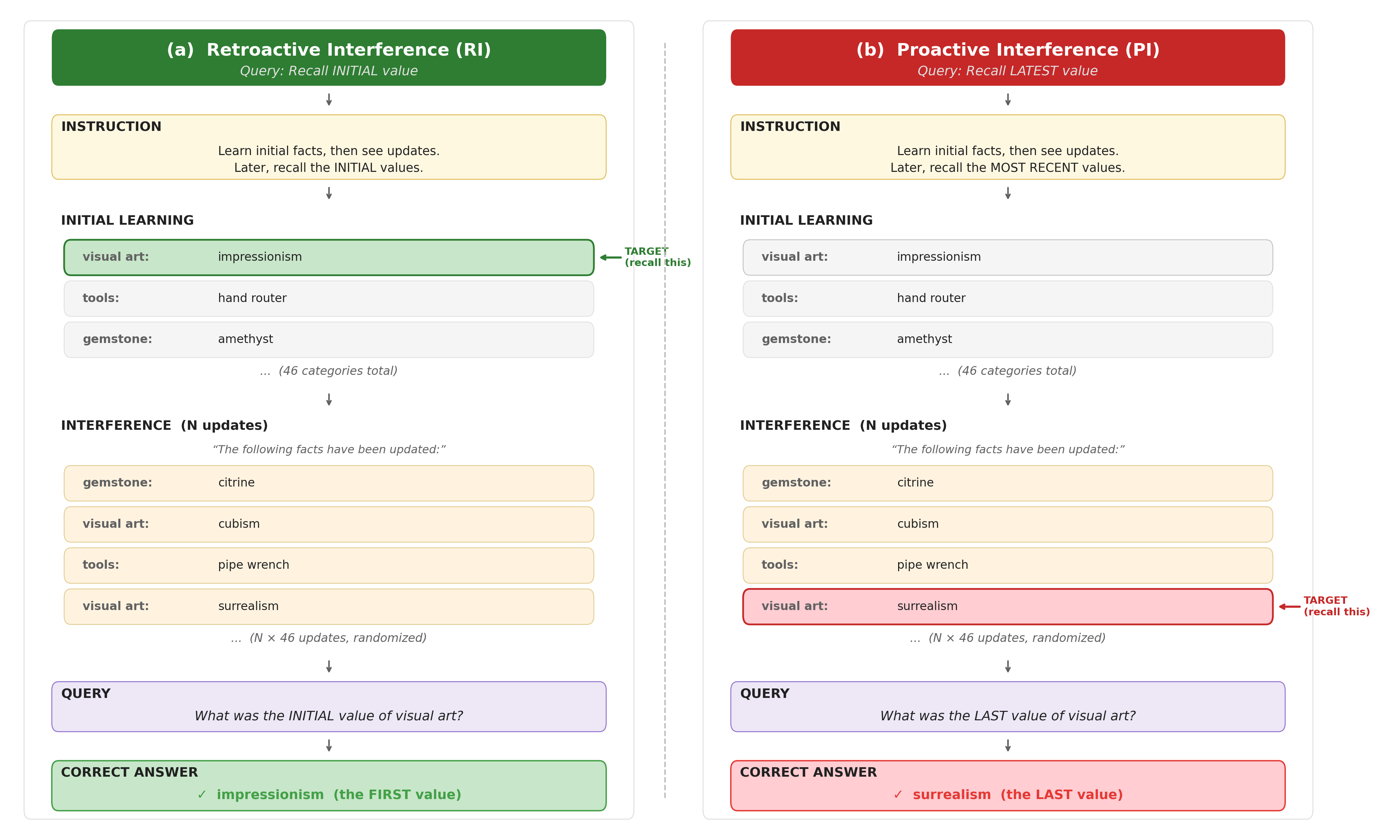}
\caption{\textbf{Shared-Stimulus Design.} Both conditions receive the same sequence of 46 initial pairs plus $N{-}1$ updates per category; RI probes the first value (green), PI the most recent (red). Identical stimuli ensure that performance differences reflect interference direction, not search difficulty.}
\label{fig:paradigm}
\end{figure}

At retrieval, the model must produce either the first value ever assigned (RI condition) or the most recently assigned value (PI condition). Both conditions receive identical input sequences; only the query target differs. This shared-stimulus design isolates the direction of interference from any confound in stimulus exposure (Figure~\ref{fig:paradigm}). All experiments fix temperature $= 0$; session-to-session variability comes from freshly sampled random sequences, not from stochastic decoding.

\paragraph{Dataset.}
Our stimulus pool spans 46 semantic categories (e.g., \emph{visual art}, \emph{musical instrument}, \emph{programming language}), each backed by 478--624 values that are genuine real-world instances of their category (e.g., ``illuminated manuscript,'' ``hand router,'' ``harpsichord'') rather than synthetic strings or numeric suffixes. This ensures that any retrieval failure reflects interference, not confusion from artificial stimuli. Each session draws a fresh random sample from this pool, so sequences vary across runs while every value remains a plausible category member. Category reassignments are shuffled so that consecutive updates never target the same key, mirroring how concurrent data streams interleave in practice. Since every query explicitly names the target category, the task minimizes search difficulty; incorrect responses therefore signal interference rather than retrieval-by-search failure.

\paragraph{Metrics.}
Because single-level accuracy captures only a snapshot, we summarize each model's full interference profile with an area-under-the-curve measure. The \textbf{Retroactive Interference Endurance Score (RIES)} integrates accuracy over a log-scaled interference axis, giving equal weight to each order of magnitude:
\begin{equation}
\text{RIES} = \int_{N=3}^{300} A_{\text{RI}}(N) \, d(\log_{10}(N + 1))
\end{equation}
Here $A_{\text{RI}}(N)$ denotes accuracy (in percentage points) on initial-value queries at interference level $N$. The theoretical maximum (100\% at every level) is RIES $\approx$ 188; across our 39 models the observed range is 113.6--186.4. The analogous \textbf{PIES} (Proactive Interference Endurance Score) applies the same computation to most-recent-value queries. When RIES exceeds PIES, the model tolerates retroactive interference better than proactive, meaning PI degrades performance more severely.

\paragraph{Models.}
We tested 54 LLMs in total; 39 produced complete data across all six interference levels and form the basis of our paired RI--PI comparison (Sections~4.1--4.5), spanning 1B to 2.5T parameters (Appendix~\ref{appendix:models}). The remaining 15 models produced partial data at low-to-moderate levels and are included only in the error analysis (Section~4.6). Models include open-weight (Llama, Qwen, DeepSeek) and proprietary (GPT-5, Claude-4.5, Gemini, o1, o3) architectures, with context windows from 8K to 2M tokens.

\paragraph{Adaptive Stopping with Bootstrap CI.}
Each session draws a \textit{fresh random sequence} from the value pool, ensuring true statistical independence despite deterministic (temperature $= 0$) generation. Sessions run until the 95\% bootstrap CI (10{,}000 resamples) narrows below 0.10 for two consecutive sessions (minimum 5 sessions). Standard models cap at 50 sessions; reasoning models (o-series, DeepSeek-R1) at 20. Sessions extracting $<$50\% of categories are excluded as parse failures.

\section{Results}

\subsection{Retroactive Interference in LLMs}

Retroactive interference is pervasive: every one of the 39 models tested shows meaningful degradation as conflicting updates accumulate. At the lowest interference level ($N = 3$), models average 93.9\% accuracy (SD=18.1\%), and this level holds through $N = 50$ (96.0\%, a ceiling effect). Beyond that threshold, performance falls steeply: 72.9\% at $N = 100$, 45.5\% at $N = 200$, and only 24.6\% at $N = 300$, a 71-point decline that demonstrates how severely subsequent updates erode access to initially learned associations.

The spread across individual models is wide (Table~\ref{tab:top_bottom}). Reasoning architectures such as o1 and o3 retain near-perfect recall even under maximum interference, whereas smaller models (Llama-3.2-1b, Nova-micro) fall to 2\%. RIES spans from 113.6 to 186.4, raising the question of what factors account for this variation.

\begin{table}[t]
\centering
\small
\caption{\textbf{Range of RI Resistance.} The five highest- and five lowest-scoring models by RIES, with accuracy at minimal ($N$=3) and maximal ($N$=300) interference.}
\label{tab:top_bottom}
\begin{tabular}{lccc@{\hskip 12pt}lccc}
\toprule
\multicolumn{4}{c}{\textit{Top 5 (Most Resistant)}} & \multicolumn{4}{c}{\textit{Bottom 5 (Least Resistant)}} \\
\cmidrule(r){1-4} \cmidrule(l){5-8}
Model & RIES & @3 & @300 & Model & RIES & @3 & @300 \\
\midrule
\rowcolor{green!6} o1 & 186.4 & 100 & 100 & Llama-3.2-3b\footnotemark & 113.6 & 0 & 76 \\
\rowcolor{green!6} o1-preview & 186.4 & 100 & 99 & Llama-3.2-1b & 126.1 & 74 & 2 \\
\rowcolor{green!6} GPT-5 & 186.4 & 100 & 38 & Nova-micro & 126.1 & 95 & 2 \\
\rowcolor{green!6} o3 & 185.8 & 100 & 100 & GPT-4.1-nano & 128.2 & 83 & 7 \\
\rowcolor{green!6} o3-mini & 183.3 & 100 & 68 & Llama-3.1-8b & 128.3 & 59 & 14 \\
\bottomrule
\end{tabular}
\footnotetext{Llama-3.2-3b produced unparseable output at $N$=3 across all sessions (0\% extracted accuracy); higher $N$ values yielded valid responses.}
\end{table}

\subsection{Model Size Correlates with RI Resistance}

Which model-level properties predict RI resistance? We regressed RIES on two candidate factors: total parameter count and declared context window length.

\begin{figure}[t]
\centering
\includegraphics[width=0.5\textwidth, trim=5 5 5 5, clip]{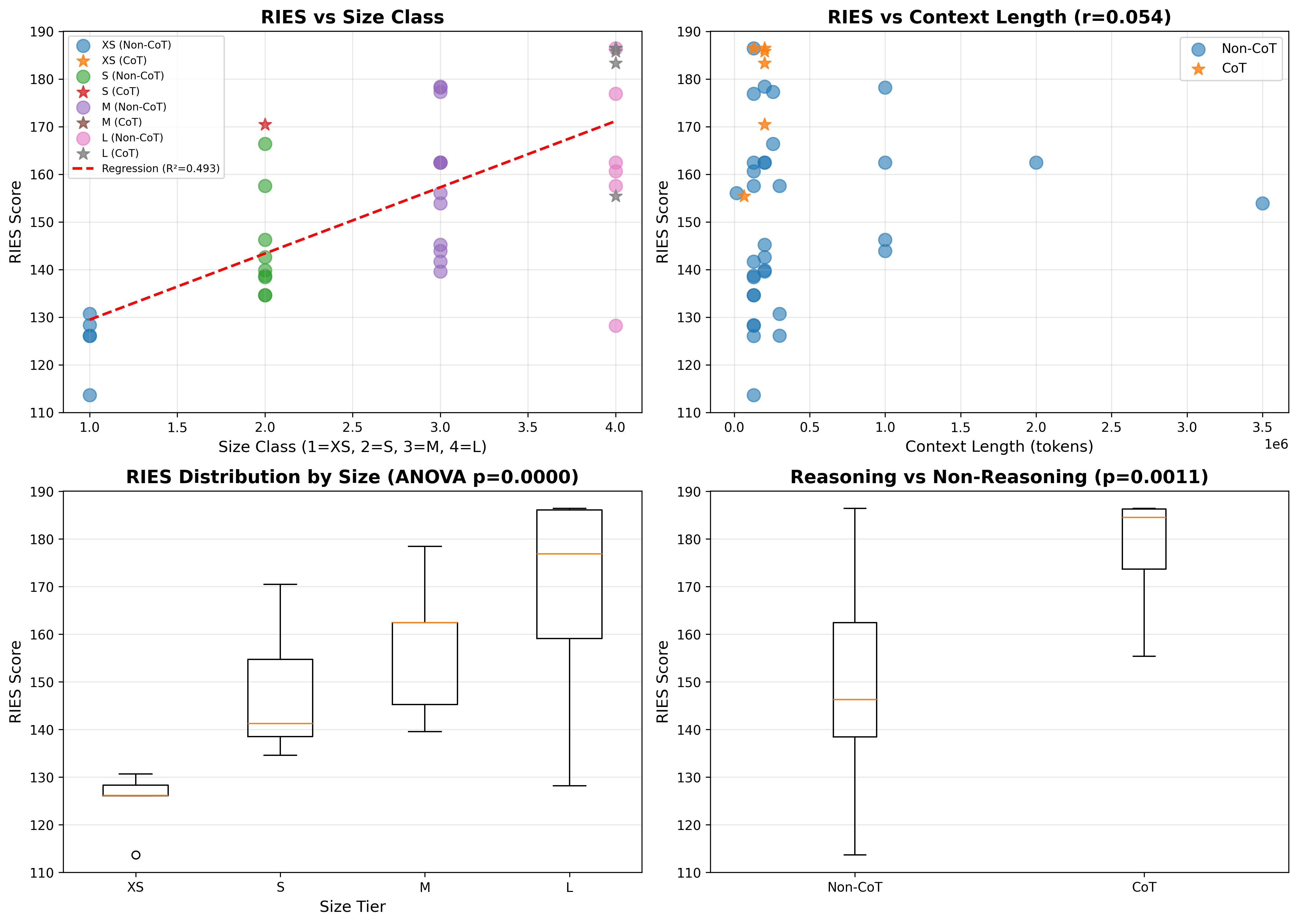}
\caption{\textbf{Scale Helps RI but Not PI.} (A) Parameter count accounts for 49\% of variance in RIES ($p < 0.0001$). (B) Context window length is uninformative ($R^2 = 0.003$, n.s.). The same regression applied to PIES yields $R^2 = 0.06$ (n.s.), confirming that model size does not predict PI resistance.}
\label{fig:regression}
\end{figure}

\textbf{Size correlates with RI resistance.} A linear regression of parameter count on RIES yields $R^2 = 0.491$ ($\beta = 0.70$, $p < 0.0001$; Figure~\ref{fig:regression}): nearly half of the between-model variance in RI resistance is attributable to scale alone. Grouping models into size tiers confirms the trend: the largest tier ($>$500B) averages RIES of 171.2 versus 142.8 for the smallest ($<$10B), a 28-point gap ($F(3, 35) = 11.83$, $p < 0.0001$).

\textbf{Context length does not.} Declared context window size bears no relationship to RIES ($R^2 = 0.003$, $p = 0.746$); in our data, a 100B-parameter model with a 32K window outperforms a 10B-parameter model with a 1M window. Interference resistance thus tracks representational capacity rather than buffer length.

\textbf{Size does not predict PI resistance.} Applying the same regression to PIES yields $R^2 = 0.06$ ($p = 0.14$, n.s.), lower than the $R^2 = 0.26$ reported by Wang and Sun~\cite{wang2025unable} on a smaller, predominantly dense model set. Our sample includes substantially more MoE architectures whose total parameter counts overstate effective capacity; this, combined with greater architectural diversity across 39 models, likely attenuates the size--PI relationship. The positive size--RI link holds whether one measures dense-only parameters ($R^2 = 0.54$), total parameters ($0.49$), or estimated active parameters ($R^2 = 0.30$, $p < 0.001$). That scale predicts RI but not PI constitutes initial evidence that the two interference types tap different underlying mechanisms.

\paragraph{Mechanistic interpretation.}
The \textit{superposition} framework \cite{elhage2022superposition} offers an intuitive account. In lower-capacity networks, feature vectors overlap, so each new encoding partially corrupts earlier ones; larger networks can assign more nearly orthogonal directions, letting initial values persist. This explains why scale alleviates RI; PI, by contrast, is bottlenecked at the attention-selection stage, an architectural property that additional parameters do not alter (see Discussion).

\subsection{Reasoning Models Excel at RI, but at a Hidden Cost}

Beyond raw scale, does extended inference-time computation confer additional protection? To find out, we separated the six chain-of-thought reasoning models \cite{wei2022chain,openai2024reasoning} (o1, o1-preview, o3, o3-mini, o4-mini, DeepSeek-R1) from the 33 standard models.

\textbf{RI benefits substantially.} Reasoning models average RIES = 178.0 (SD=12.6) versus 150.3 (SD=18.2) for standard models, an 18.4\% advantage ($t(37) = 3.54$, $p = 0.001$, $d = 1.76$). Five of the six fall in the overall top 10, consistent with the idea that deliberative processing reinforces the consolidation of initial encodings.

\textbf{PI does not.} The same models that dominate the RI ranking fare poorly on PI. The starkest case is o1: first in RIES (186.4) yet 35th in PIES (22.0), an 8.5$\times$ disparity within a single model. This within-model dissociation underscores that consolidation and recency retrieval are separable: reasoning strengthens the former while leaving the latter unchanged or worse.

\textbf{Practical implication:} Reasoning architectures suit applications requiring reliable historical recall (document review, knowledge retrieval) but may be a poor fit for tasks centered on tracking evolving state (multi-turn dialogue, live monitoring).

\subsection{Decay Patterns Reveal Architectural Diversity}

\begin{figure}[t]
\centering
\includegraphics[width=0.48\textwidth]{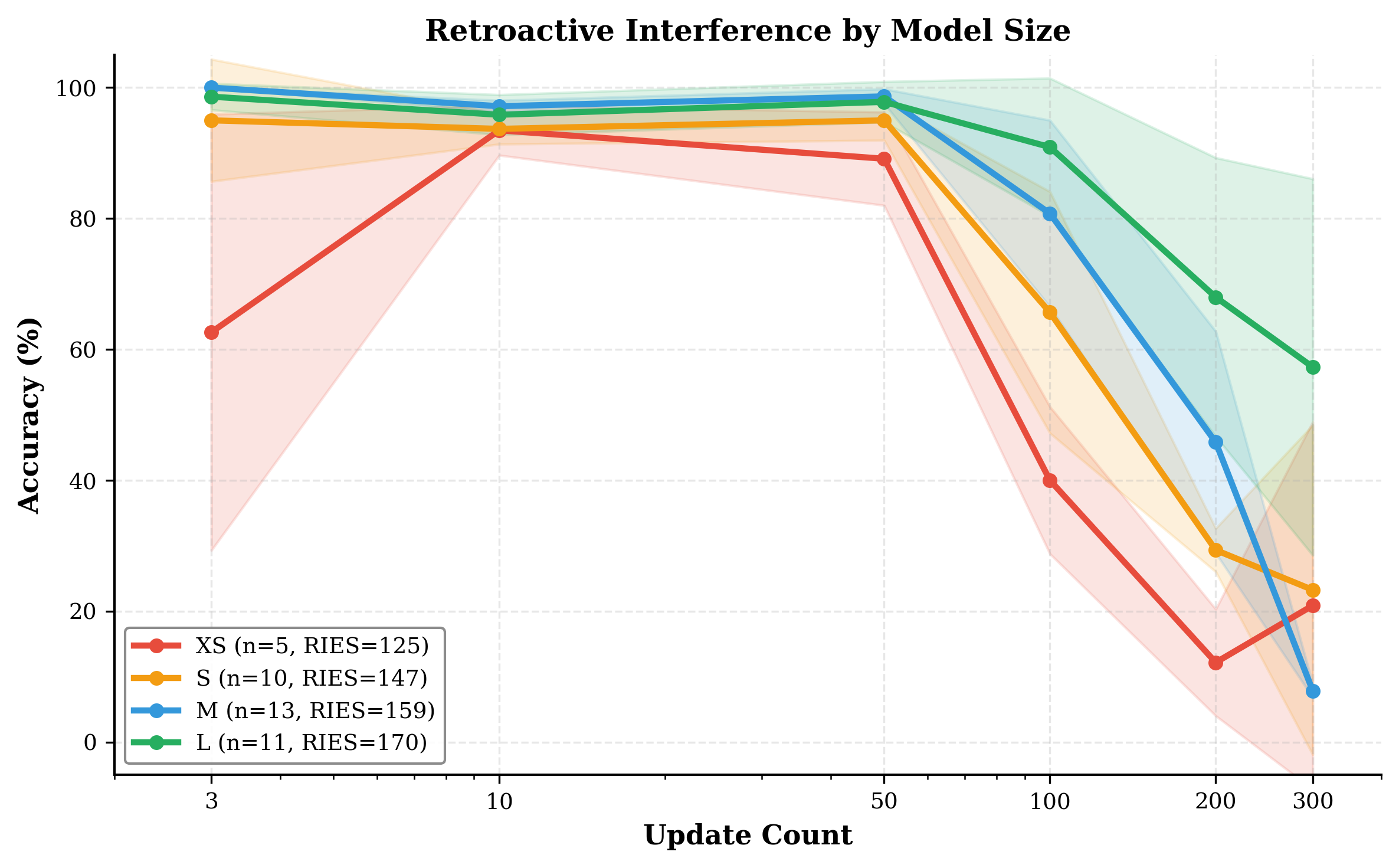}
\hfill
\includegraphics[width=0.48\textwidth]{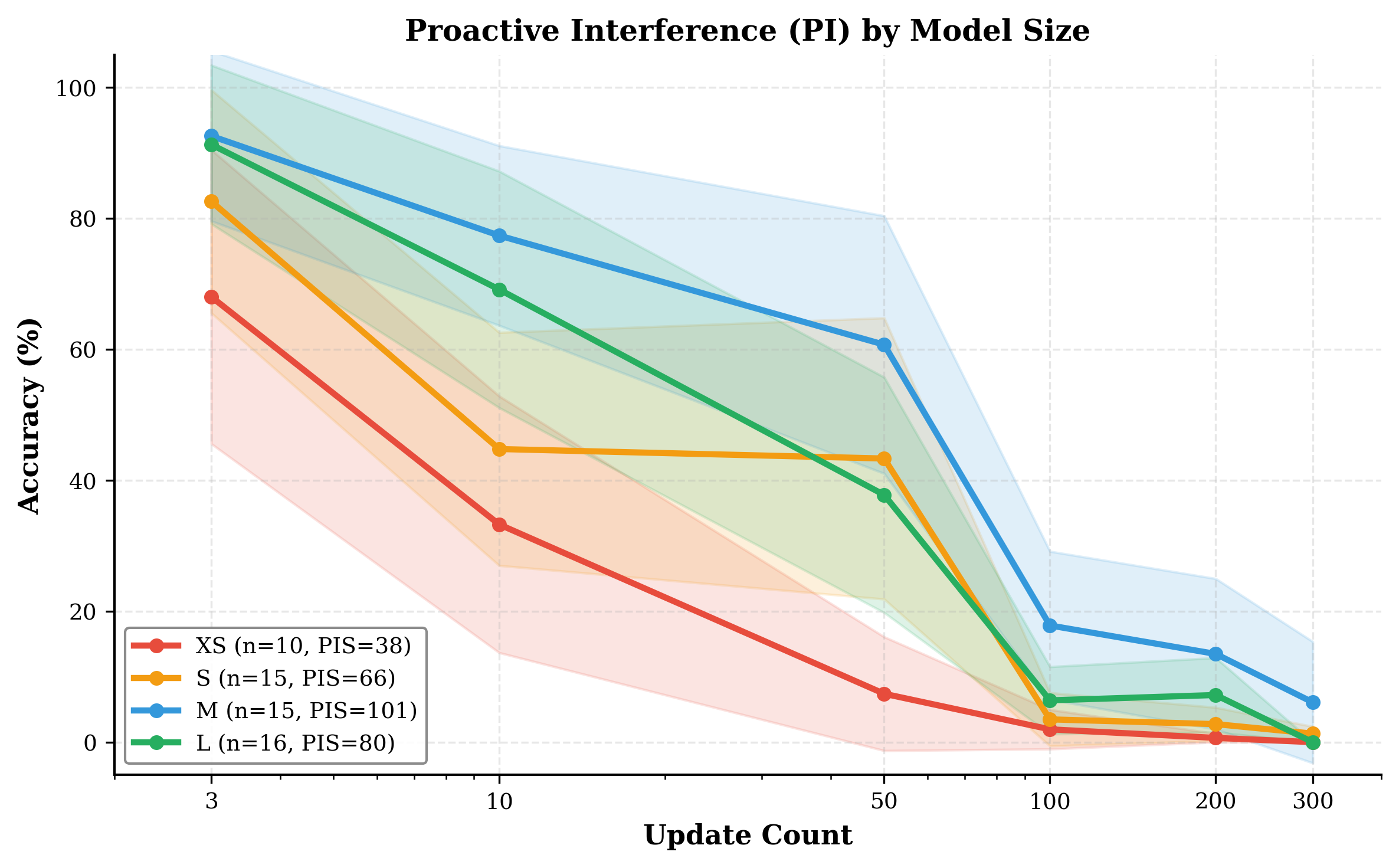}
\caption{\textbf{Decay Curves by Model Size: RI (left) vs PI (right).} Both plots use the same 39 models with complete paired data. RI shows clear size stratification (L tier $>$50\% at $N$=300, XS $<$25\%). PI shows no such ordering: the M tier outperforms the larger L tier, confirming that size predicts RI ($R^2 = 0.49$) but not PI ($R^2 = 0.06$).}
\label{fig:decay_size}
\end{figure}

PI accuracy follows a near-universal log-linear trajectory across architectures, consistent with Wang and Sun~\cite{wang2025unable}. RI decay tells a different story (Figure~\ref{fig:decay_size}): the best-fitting function varies: exponential for 35\% of models (steep early loss), polynomial for 30\% (gradual erosion), log-quadratic for 26\% (plateau then collapse), and power-law for 9\% (O-series reasoning models). This uniformity in PI versus heterogeneity in RI suggests that PI operates through shared attention-selection mechanisms, while RI depends on architecture-specific consolidation processes. Session-level statistics reinforce the split: RI converges within 5--10 sessions across all $N$, whereas PI at $N \geq 200$ is bimodal, alternating between near-ceiling ($>$85\%) and near-floor ($<$5\%).

\subsection{RI and PI Engage Distinct Mechanisms}

A single shared ``memory capacity'' would predict that models excelling at RI should also excel at PI. We tested this by correlating RIES and PIES across all 39 models.

\begin{figure}[t]
\centering
\includegraphics[width=0.75\textwidth, trim=5 5 5 5, clip]{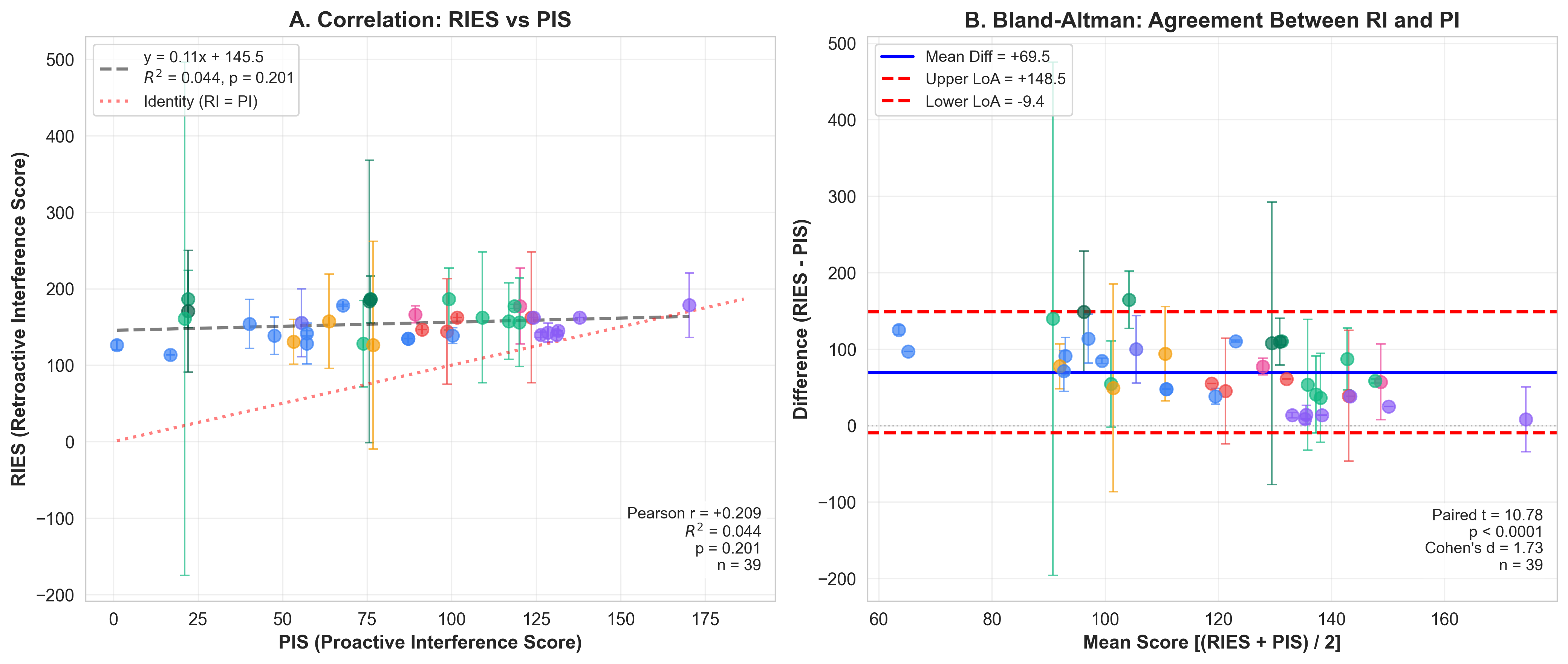}
\caption{\textbf{Dissociation Between RI and PI Resistance.} (A) RIES vs.\ PIES across 39 models: the correlation is weak ($r = 0.21$, 95\% CI $[-0.11, 0.49]$, $p = 0.20$) and every model falls above the identity line (RIES $>$ PIES throughout). (B) Bland-Altman analysis shows a systematic offset of +69.5 points (95\% limits of agreement: $-9.4$ to $+148.5$), confirming that the two scores capture different aspects of model performance.}
\label{fig:correlation}
\end{figure}

\textbf{The two scores are largely independent.} Pearson $r = 0.21$ ($R^2 = 0.044$, $p = 0.201$, 95\% CI $= [-0.11, 0.49]$; Figure~\ref{fig:correlation}). The confidence interval spans zero and shared variance is under 5\%, so a model's RI score carries little predictive value for its PI score. This is difficult to reconcile with a unified-capacity account.

\textbf{PI is universally harder.} Every model without exception (39/39) achieves higher RIES than PIES. The mean gap is 69.5 points (paired $t(38) = 10.71$, $p < 0.0001$, Cohen's $d = 1.73$), indicating that proactive interference poses a categorically greater challenge for current LLMs.

\textbf{The reverse of the human pattern.} In human paired-associate paradigms, RI generally produces more forgetting than PI \cite{macleod2024interference,underwood1957interference}. Transformers invert this, exhibiting what we term ``primacy protection.'' The degree of inversion varies: o1 shows an 8.5$\times$ gap (RIES = 186.4, PIES = 22.0), GPT-5 a 1.9$\times$ gap, and Claude-4.5-opus only 1.05$\times$ (RIES = 178.4, PIES = 170.3), suggesting that some designs achieve a more balanced profile.

\subsection{Error Analysis Reveals Distinct Failure Modes}

We examined 5,409 RI errors and 9,753 PI errors across all 54 models (the 15 with partial data contributed errors at whichever levels they completed). An algorithmic classifier compared each incorrect response against the full stimulus history and assigned it to one of five categories: \textit{retrieval failure} (no value produced), \textit{same-key intrusion} (correct category, wrong position), \textit{cross-key intrusion} (different category), \textit{hallucination} (value absent from the sequence), or \textit{partial match} (near-correct but inexact).

\textbf{The two conditions produce qualitatively different error profiles} (Table~\ref{tab:errors}). RI is dominated by retrieval failures (50.8\%), where the model returns nothing, as if the initial encoding were overwritten. PI is dominated by same-key intrusions (56.1\%), where the model confidently produces an earlier value from the correct category, indicating active competition from prior encodings.

\textbf{Fabrication is rare in both conditions.} Under a strict criterion (values entirely absent from the stimulus sequence), hallucination accounts for only 0.9\% of RI errors and 0.2\% of PI errors. Same-key intrusions, which involve values that \textit{were} presented but at the wrong position, are positional errors, not hallucinations. Models confuse the ordering of seen material rather than invent new material, situating interference as a selection problem rather than a generation problem.

\textbf{Self-monitoring surfaces only under PI.} An LLM-as-judge analysis finds that 8.6\% of PI retrieval failures contain explicit refusals, versus 0\% for RI. Extending the self-knowledge findings of Kadavath et al.~\cite{kadavath2022language} and Yin et al.~\cite{yin2023large}, this shows that metacognitive signaling is interference-type-specific: competing attention candidates under PI can trigger uncertainty, whereas the diluted signal under RI is returned with unwarranted confidence.

\textbf{Positional bias confirms opposing dynamics.} At low interference the contrast is starkest: RI intrusions originate from late positions (mean normalized position 0.72 at $N = 3$), whereas PI intrusions skew early (0.12 at $N = 3$), with 14\% returning the very first value. As $N$ increases, both converge toward the midpoint (0.51 for RI, Figure~\ref{fig:position_bias}). This mirror-image directionality provides direct evidence that the two interference types recruit opposite retrieval biases.

\begin{table}[t]
\centering
\small
\caption{\textbf{How RI and PI Errors Differ.}}
\label{tab:errors}
\begin{tabular}{lcc}
\toprule
Error Type & RI (\%) & PI (\%) \\
\midrule
Retrieval failure & \cellcolor{blue!8}\textbf{50.8} & 42.5 \\
Same-key intrusion & 46.0 & \cellcolor{red!8}\textbf{56.1} \\
Cross-key intrusion & 0.6 & 0.7 \\
Hallucination & 0.9 & 0.2 \\
Partial match & 1.8 & 0.7 \\
\bottomrule
\end{tabular}
\end{table}

\section{Discussion}

\paragraph{Evidence Consistent with Dual-Process Memory.}
Four independent results converge on the conclusion that RI and PI are not manifestations of a single capacity limit. (i) The two endurance scores correlate only weakly ($r = 0.21$). (ii) Parameter count explains substantial variance in RIES ($R^2 = 0.49$) but negligible variance in PIES ($R^2 = 0.06$). (iii) Error profiles diverge qualitatively: RI produces passive retrieval failures, whereas PI produces active intrusions from earlier positions, a pattern reminiscent of retrieval-induced forgetting in humans \cite{anderson1994remembering}. (iv) RI accuracy converges reliably within 5--10 sessions, while PI at high interference ($N \geq 200$) alternates bimodally between near-ceiling ($>85\%$) and near-complete collapse ($<5\%$), consistent with a winner-take-all competition in which softmax attention must commit to either primacy or recency.

This pattern maps onto the consolidation--retrieval distinction from cognitive science \cite{wixted2004psychology}: RI probes whether initial encodings survive overwriting (a capacity-bound process), while PI probes whether attention can select recent information over entrenched competitors (an architecture-bound process).

\paragraph{Superposition May Explain the Capacity Dependence.}
The superposition hypothesis \cite{elhage2022superposition} offers one mechanistic account for why scale helps RI but not PI. When a network has fewer dimensions than concepts to represent, features must share directions in activation space \cite{scherlis2022polysemanticity}; each new encoding therefore partially overwrites earlier ones. Adding parameters provides more nearly orthogonal slots, allowing initial values to persist alongside subsequent updates. This account predicts exactly the asymmetry we observe: scale expands representational room (aiding RI), but does not change the attention-selection bottleneck that governs PI.

\paragraph{Why Transformers Differ from Humans.}
Under experimental conditions most comparable to ours, human participants typically forget old associations more readily than new ones (RI $>$ PI) \cite{macleod2024interference,underwood1957interference}. Transformers exhibit the reverse. A natural explanation lies in the mechanics of causal self-attention: because each early token is attended to by every later token, while later tokens attend only backward, early positions accumulate disproportionate influence: the ``attention sink'' phenomenon \cite{xiao2024efficient} and the broader ``lost in the middle'' pattern \cite{liu2024lost}. The resulting primacy bias shields initial encodings at the expense of recent ones.

\paragraph{Practical Implications.}
One actionable takeaway is that \textbf{representational capacity (parameter count) outweighs buffer size (context length)} as a predictor of interference resistance. Table~\ref{tab:model_selection} collects the main empirical patterns; we stress that these emerge from a single controlled paradigm and should be validated in application-specific settings before informing deployment decisions.

\begin{table}[t]
\centering
\small
\caption{\textbf{Summary of Key Empirical Patterns.} Derived from a single paradigm; application-specific validation is needed before generalization.}
\label{tab:model_selection}
\setlength{\tabcolsep}{5pt}
\begin{tabular}{>{\raggedright\arraybackslash}p{2.4cm}>{\centering\arraybackslash}p{2.8cm}>{\centering\arraybackslash}p{2.8cm}>{\raggedright\arraybackslash}p{2.8cm}}
\toprule
 & \cellcolor{green!10}\textbf{RI Resistance} & \cellcolor{red!8}\textbf{PI Resistance} & \textbf{Key Insight} \\
\midrule
\rowcolor{gray!5} Reasoning models & \cellcolor{green!12}High (+18\%) & \cellcolor{red!12}Lowest & Consolidation $\neq$ retrieval \\[3pt]
Large dense & \cellcolor{green!8}Strong & \cellcolor{green!8}Strong & Balanced interference profile \\[3pt]
\rowcolor{gray!5} Parameter count & \cellcolor{green!12}Predictive & \cellcolor{red!12}Not predictive & Capacity helps RI only \\[3pt]
Context length & \cellcolor{red!8}No effect & \cellcolor{red!8}No effect & Buffer size irrelevant \\
\bottomrule
\end{tabular}
\end{table}

\section{Conclusion}

This study offers the first controlled comparison of retroactive and proactive interference within a single LLM evaluation framework. Across every model tested, proactive interference proved substantially more damaging than retroactive ($d = 1.73$, 39/39 models), the inverse of the pattern typically reported for human paired-associate learning. Four converging results point to computationally distinct processes behind the two phenomena: weak correlation between RI and PI resistance coupled with divergent scaling behavior; qualitatively different error signatures (passive retrieval failure for RI, active primacy intrusion for PI, hallucination below 1\% in both); and contrasting session-level stability, with RI decaying smoothly while PI collapses bimodally at high interference loads.

Critically, these failures are largely invisible to end users: more than 97\% of incorrect retrievals carry no uncertainty signal. The asymmetry extends even to self-knowledge: 8.6\% of PI retrieval failures elicit an explicit refusal, whereas RI failures are entirely silent (0\% refusal). Taken together, these findings indicate that transformer attention inherently shields early encodings at the expense of recency, producing confident yet incorrect outputs when in-context information evolves.

\paragraph{Future Directions.}
Several avenues warrant further work. First, \textit{ecologically grounded tasks} such as evolving narratives \cite{kocisky2018narrativeqa} or semi-structured records (clinical logs, legal filings) with verifiable updates should complement our controlled paradigm. Second, \textit{mechanistic probing} via attention-head ablation and training-checkpoint comparisons \cite{biderman2023pythia} on open-weight models could clarify whether the PI $>$ RI asymmetry is architectural or learned. Third, \textit{mitigation strategies} such as recency-aware attention re-weighting, deliberate context ordering, or positional debiasing \cite{liu2024lost} deserve evaluation as means of reducing PI vulnerability without sacrificing RI robustness.

\section*{Limitations}

All results rest on a single paradigm (AB-AC paired associates), providing tight control but no guarantee of transfer to other interference manipulations or naturalistic text. Our mechanistic accounts (superposition, attention sinks, winner-take-all) are post-hoc and unconfirmed by attention probing; a single-mechanism alternative (two bottlenecks of one process) cannot be ruled out from behavioral data alone. Further caveats: (1) proprietary parameter counts are estimates; (2) $n = 39$ limits power for moderate RI--PI correlations, though four converging lines of evidence support the dissociation; (3) only transformer architectures were tested; (4) English-only stimuli; (5) temperature $= 0$ throughout; (6) non-converging PI conditions carry wide CIs reflecting genuine bimodality; (7) paired RI--PI analyses use 39 models with complete data; error analysis uses all 54 (including 15 partial) to maximize statistical power for characterizing failure modes.

\section*{Ethical Considerations}

All experiments rely on synthetic category-value stimuli; no human participants were recruited and no personal data were collected. Our characterization of model-specific interference vulnerabilities is intended to support informed deployment decisions. We acknowledge a dual-use dimension: knowledge of how interference degrades recall could, in principle, be used to craft adversarial inputs. We believe the practical benefit, helping practitioners choose appropriate models and anticipate failure modes, substantially outweighs this risk.

\bibliographystyle{splncs03}
\bibliography{references}

@article{brown2020language,
  title={Language models are few-shot learners},
  author={Brown, Tom and Mann, Benjamin and Ryder, Nick and Subbiah, Melanie and Kaplan, Jared D and Dhariwal, Prafulla and Neelakantan, Arvind and Shyam, Pranav and Sastry, Girish and Askell, Amanda and others},
  journal={Advances in neural information processing systems},
  volume={33},
  pages={1877--1901},
  year={2020},
  url={https://arxiv.org/abs/2005.14165}
}

@article{achiam2023gpt4,
  title={GPT-4 technical report},
  author={Achiam, Josh and Adler, Steven and Agarwal, Sandhini and Ahmad, Lama and Akkaya, Ilge and Aleman, Florencia Leoni and Almeida, Diogo and Altenschmidt, Janko and Altman, Sam and Anadkat, Shyamal and others},
  journal={arXiv preprint arXiv:2303.08774},
  year={2023},
  url={https://arxiv.org/abs/2303.08774}
}

@inproceedings{wang2025unable,
  title={Unable to forget: Proactive interference reveals working memory limits in LLMs beyond context length},
  author={Wang, Chupei and Sun, Jiaqiu Vince},
  booktitle={ICML 2025 Workshop on Long Context Foundation Models},
  year={2025},
  url={https://arxiv.org/abs/2506.08184}
}

@article{jenkins1924obliviscence,
  title={Obliviscence during sleep and waking},
  author={Jenkins, John G and Dallenbach, Karl M},
  journal={The American journal of psychology},
  volume={35},
  number={4},
  pages={605--612},
  year={1924},
  publisher={JSTOR},
  url={https://www.semanticscholar.org/paper/Obliviscence-During-Sleep-and-Waking.-Jenkins-Dallenbach/b4fe2e847e01a1e775296d1863364be11b49a796}
}

@article{underwood1957interference,
  title={Interference and forgetting},
  author={Underwood, Benton J},
  journal={Psychological review},
  volume={64},
  number={1},
  pages={49},
  year={1957},
  publisher={American Psychological Association},
  url={https://psycnet.apa.org/record/1958-00251-001}
}

@article{wixted2004psychology,
  title={The psychology and neuroscience of forgetting},
  author={Wixted, John T},
  journal={Annual review of psychology},
  volume={55},
  pages={235--269},
  year={2004},
  publisher={Annual Reviews},
  url={https://pubmed.ncbi.nlm.nih.gov/14744216/}
}

@article{wei2022chain,
  title={Chain-of-thought prompting elicits reasoning in large language models},
  author={Wei, Jason and Wang, Xuezhi and Schuurmans, Dale and Bosma, Maarten and Xia, Fei and Chi, Ed and Le, Quoc V and Zhou, Denny and others},
  journal={Advances in Neural Information Processing Systems},
  volume={35},
  pages={24824--24837},
  year={2022},
  url={https://arxiv.org/abs/2201.11903}
}

@techreport{openai2024reasoning,
  title={Learning to reason with LLMs},
  author={{OpenAI}},
  institution={OpenAI},
  year={2024},
  url={https://openai.com/index/learning-to-reason-with-llms/}
}

@article{engle2002working,
  title={Working memory capacity as executive attention},
  author={Engle, Randall W},
  journal={Current directions in psychological science},
  volume={11},
  number={1},
  pages={19--23},
  year={2002},
  publisher={SAGE Publications},
  url={https://journals.sagepub.com/doi/10.1111/1467-8721.00160}
}

@article{mccloskey1989catastrophic,
  title={Catastrophic interference in connectionist networks: The sequential learning problem},
  author={McCloskey, Michael and Cohen, Neal J},
  journal={Psychology of learning and motivation},
  volume={24},
  pages={109--165},
  year={1989},
  publisher={Elsevier},
  url={https://www.sciencedirect.com/science/chapter/bookseries/abs/pii/S0079742108605368}
}

@article{french1999catastrophic,
  title={Catastrophic forgetting in connectionist networks},
  author={French, Robert M},
  journal={Trends in cognitive sciences},
  volume={3},
  number={4},
  pages={128--135},
  year={1999},
  publisher={Elsevier},
  url={https://pubmed.ncbi.nlm.nih.gov/10322466/}
}

@article{kaplan2020scaling,
  title={Scaling laws for neural language models},
  author={Kaplan, Jared and McCandlish, Sam and Henighan, Tom and Brown, Tom B and Chess, Benjamin and Child, Rewon and Gray, Scott and Radford, Alec and Wu, Jeffrey and Amodei, Dario},
  journal={arXiv preprint arXiv:2001.08361},
  year={2020},
  url={https://arxiv.org/abs/2001.08361}
}

@article{hoffmann2022chinchilla,
  title={Training compute-optimal large language models},
  author={Hoffmann, Jordan and Borgeaud, Sebastian and Mensch, Arthur and Buchatskaya, Elena and Cai, Trevor and Rutherford, Eliza and Casas, Diego de Las and Hendricks, Lisa Anne and Welbl, Johannes and Clark, Aidan and others},
  journal={arXiv preprint arXiv:2203.15556},
  year={2022},
  url={https://arxiv.org/abs/2203.15556}
}

@article{liu2024lost,
  title={Lost in the middle: How language models use long contexts},
  author={Liu, Nelson F and Lin, Kevin and Hewitt, John and Paranjape, Ashwin and Bevilacqua, Michele and Petroni, Fabio and Liang, Percy},
  journal={Transactions of the Association for Computational Linguistics},
  volume={12},
  pages={157--173},
  year={2024},
  url={https://aclanthology.org/2024.tacl-1.9.pdf}
}

@misc{kamradt2023needle,
  title={Needle In A Haystack - Pressure Testing LLMs},
  author={Kamradt, Greg},
  year={2023},
  url={https://github.com/gkamradt/LLMTest_NeedleInAHaystack}
}

@article{bai2024longbench,
  title={LongBench: A bilingual, multitask benchmark for long context understanding},
  author={Bai, Yushi and Lv, Xin and Zhang, Jiajie and Lyu, Hongchang and Tang, Jiankai and Huang, Zhidian and Du, Zhengxiao and Liu, Xiao and Zeng, Aohan and Hou, Lei and others},
  journal={arXiv preprint arXiv:2308.14508},
  year={2024},
  url={https://arxiv.org/abs/2308.14508}
}

@inproceedings{modarressi2025nolima,
  title={NoLiMa: Long-Context Evaluation Beyond Literal Matching},
  author={Modarressi, Ali and Deilamsalehy, Hanieh and Dernoncourt, Franck and Bui, Trung and Rossi, Ryan A. and Yoon, Seunghyun and Sch{\"u}tze, Hinrich},
  booktitle={Proceedings of the 42nd International Conference on Machine Learning},
  pages={44554--44570},
  year={2025},
  url={https://arxiv.org/abs/2502.05167}
}

@article{fu2025absencebench,
  title={AbsenceBench: Language Models Can't Tell What's Missing},
  author={Fu, Harvey Yiyun and Shrivastava, Aryan and Moore, Jared and West, Peter and Tan, Chenhao and Holtzman, Ari},
  journal={arXiv preprint arXiv:2506.11440},
  year={2025},
  url={https://arxiv.org/abs/2506.11440}
}

@article{ling2025longreason,
  title={LongReason: A Synthetic Long-Context Reasoning Benchmark via Context Expansion},
  author={Ling, Zhan and Liu, Kang and Yan, Kai and Yang, Yifan and Lin, Weijian and Fan, Ting-Han and Shen, Lingfeng and Du, Zhengyin and Chen, Jiecao},
  journal={arXiv preprint arXiv:2501.15089},
  year={2025},
  url={https://arxiv.org/abs/2501.15089}
}

@article{elhage2022superposition,
  title={Toy models of superposition},
  author={Elhage, Nelson and Hume, Tristan and Olsson, Catherine and Schiefer, Nicholas and Henighan, Tom and Kravec, Shauna and Hatfield-Dodds, Zac and Lasenby, Robert and Drain, Dawn and Chen, Carol and others},
  journal={Transformer Circuits Thread},
  year={2022},
  url={https://arxiv.org/abs/2209.10652}
}

@article{scherlis2022polysemanticity,
  title={Polysemanticity and capacity in neural networks},
  author={Scherlis, Adam and Sachan, Kshitij and Jermyn, Adam S and Benton, Joe and Shlegeris, Buck},
  journal={arXiv preprint arXiv:2210.01892},
  year={2022},
  url={https://arxiv.org/abs/2210.01892}
}

@inproceedings{xiao2024efficient,
  title={Efficient streaming language models with attention sinks},
  author={Xiao, Guangxuan and Tian, Yuandong and Chen, Beidi and Han, Song and Lewis, Mike},
  booktitle={International Conference on Learning Representations},
  year={2024},
  url={https://arxiv.org/abs/2309.17453}
}

@incollection{macleod2024interference,
  title={Interference theory: History and current status},
  author={MacLeod, Colin M},
  booktitle={The Oxford Handbook of Human Memory},
  pages={1173--1208},
  year={2024},
  publisher={Oxford University Press},
  url={https://www.researchgate.net/publication/382372232_Interference_Theory_History_and_Current_Status}
}

@article{wickens1970encoding,
  title={Encoding categories of words: An empirical approach to meaning},
  author={Wickens, Delos D},
  journal={Psychological Review},
  volume={77},
  number={1},
  pages={1--15},
  year={1970},
  publisher={American Psychological Association},
  url={https://www.researchgate.net/publication/232587396_Encoding_categories_of_words_An_empirical_approach_to_meaning}
}

@article{kane2001working,
  title={Working-memory capacity, proactive interference, and divided attention: Limits on long-term memory retrieval},
  author={Kane, Michael J and Engle, Randall W},
  journal={Journal of Experimental Psychology: Learning, Memory, and Cognition},
  volume={27},
  number={2},
  pages={336--358},
  year={2001},
  publisher={American Psychological Association},
  url={https://pubmed.ncbi.nlm.nih.gov/10764100/}
}

@article{biderman2023pythia,
  title={Pythia: A suite for analyzing large language models across training and scaling},
  author={Biderman, Stella and Schoelkopf, Hailey and Anthony, Quentin Gregory and Bradley, Herbie and O'Brien, Kyle and Hallahan, Eric and Khan, Mohammad Aflah and Purohit, Shivanshu and Prashanth, USVSN Sai and Raff, Edward and others},
  journal={International Conference on Machine Learning},
  pages={2397--2430},
  year={2023},
  organization={PMLR},
  url={https://arxiv.org/abs/2304.01373}
}

@article{kocisky2018narrativeqa,
  title={The {NarrativeQA} reading comprehension challenge},
  author={Ko{\v{c}}isk{\'y}, Tom{\'a}{\v{s}} and Schwarz, Jonathan and Blunsom, Phil and Dyer, Chris and Hermann, Karl Moritz and Melis, G{\'a}bor and Grefenstette, Edward},
  journal={Transactions of the Association for Computational Linguistics},
  volume={6},
  pages={317--328},
  year={2018},
  url={https://aclanthology.org/Q18-1023.pdf}
}

@article{murdock1962serial,
  title={The serial position effect of free recall},
  author={Murdock, Bennet B},
  journal={Journal of Experimental Psychology},
  volume={64},
  number={5},
  pages={482--488},
  year={1962},
  publisher={American Psychological Association},
  url={https://psycnet.apa.org/record/1963-06156-001}
}

@article{anderson1994remembering,
  title={Remembering can cause forgetting: Retrieval dynamics in long-term memory},
  author={Anderson, Michael C and Bjork, Robert A and Bjork, Elizabeth L},
  journal={Journal of Experimental Psychology: Learning, Memory, and Cognition},
  volume={20},
  number={5},
  pages={1063--1087},
  year={1994},
  publisher={American Psychological Association},
  url={https://pubmed.ncbi.nlm.nih.gov/7931095/}
}

@article{barnes1959fate,
  title={Fate of first-list associations in transfer theory},
  author={Barnes, Jean M and Underwood, Benton J},
  journal={Journal of Experimental Psychology},
  volume={58},
  number={2},
  pages={97--105},
  year={1959},
  publisher={American Psychological Association},
  url={https://psycnet.apa.org/record/1960-03903-001}
}

@article{kadavath2022language,
  title={Language models (mostly) know what they know},
  author={Kadavath, Saurav and Conerly, Tom and Askell, Amanda and Henighan, Tom and Drain, Dawn and Perez, Ethan and Schiefer, Nicholas and Hatfield-Dodds, Zac and DasSarma, Nova and Tran-Johnson, Eli and others},
  journal={arXiv preprint arXiv:2207.05221},
  year={2022},
  url={https://arxiv.org/abs/2207.05221}
}

@inproceedings{yin2023large,
  title={Do large language models know what they don't know?},
  author={Yin, Zhangyue and Sun, Qiushi and Guo, Qipeng and Wu, Jiawen and Qiu, Xipeng and Huang, Xuanjing},
  booktitle={Findings of the Association for Computational Linguistics: ACL 2023},
  year={2023},
  url={https://aclanthology.org/2023.findings-acl.551.pdf}
}

\appendix

\section{Model Specifications}
\label{appendix:models}

\vspace{-4pt}
Table~\ref{tab:models} lists all 39 models sorted by parameter count. We report total parameters; only models using chain-of-thought by default (o-series, DeepSeek-R1) are classified as reasoning.

\begin{table}[H]
\centering
\scriptsize
\setlength{\tabcolsep}{2pt}
\caption{\textbf{All 39 Models with RIES and PIES Scores.} Sorted by parameter count (B = billions). A = architecture (D = Dense, M = MoE). \textsuperscript{R} = reasoning model. Cl = Claude, Gem = Gemini, DS = DeepSeek.}
\label{tab:models}
\begin{tabular}{lrccc@{\hskip 8pt}lrccc}
\toprule
\textbf{Model} & \textbf{B} & \textbf{A} & \textbf{RIES} & \textbf{PIES} &
\textbf{Model} & \textbf{B} & \textbf{A} & \textbf{RIES} & \textbf{PIES} \\
\midrule
GPT-5            & 2500 & M & 186.4 & 99.2  & GPT-3.5-turbo    & 175 & M & 156.1 & 120.0 \\
GPT-5-mini       & 2500 & M & 162.5 & 109.1 & Llama-4-scout    & 109 & M & 153.9 & 40.2  \\
GPT-5-nano       & 2500 & M & 160.6 & 21.0  & Gem-2.0-flash    & 100 & M & 162.5 & 101.6 \\
o3\textsuperscript{R}  & 2000 & M & 185.8 & 76.0  & Gem-2.0-fl-lite  & 100 & M & 143.9 & 98.6  \\
o3-mini\textsuperscript{R} & 2000 & M & 183.3 & 75.6  & Llama-3.2-90b   & 90  & D & 134.6 & 87.1  \\
o1\textsuperscript{R}  & 1760 & M & 186.4 & 22.0  & Nova-pro         & 70  & D & 157.5 & 63.7  \\
o1-preview\textsuperscript{R} & 1760 & M & 186.4 & 76.0  & Llama-3.1-70b   & 70  & D & 134.6 & 87.1  \\
GPT-4.1          & 1760 & M & 176.9 & 118.6 & Llama-3.3-70b    & 70  & D & 138.7 & 100.3 \\
GPT-4.1-mini     & 1760 & M & 157.6 & 116.9 & Gem-2.5-fl-lite  & 40  & M & 146.3 & 91.3  \\
GPT-4.1-nano     & 1760 & M & 128.2 & 73.9  & o4-mini\textsuperscript{R} & 30 & M & 170.5 & 22.0  \\
DS-R1\textsuperscript{R} & 671 & D & 155.4 & 55.6  & Qwen-3-cod-30b  & 30  & M & 166.4 & 89.3  \\
Llama-3.1-405b   & 405  & D & 141.7 & 57.2  & Cl-4.5-haiku     & 15  & D & 142.6 & 128.5 \\
Llama-4-mav      & 400  & M & 178.3 & 67.9  & Llama-3.2-11b    & 11  & D & 138.4 & 47.6  \\
Cl-4.5-opus      & 300  & D & 178.4 & 170.3 & Cl-3.5-haiku     & 10  & D & 139.8 & 126.4 \\
Gem-2.5-pro      & 300  & M & 162.5 & 123.6 & Llama-3.1-8b     & 8   & D & 128.3 & 57.1  \\
Cl-4.5-sonnet    & 250  & D & 162.5 & 124.2 & Nova-lite         & 7   & D & 130.7 & 53.2  \\
Qwen-3-VL-235b   & 235  & M & 177.3 & 120.2 & Llama-3.2-3b     & 3   & D & 113.6 & 16.8  \\
Cl-4-opus        & 200  & D & 139.6 & 131.0 & Nova-micro        & 1   & D & 126.1 & 76.8  \\
Cl-4-sonnet      & 200  & D & 162.5 & 137.8 & Llama-3.2-1b     & 1   & D & 126.1 & 1.0   \\
Cl-3-opus        & 175  & D & 145.2 & 131.5 & & & & & \\
\bottomrule
\end{tabular}
\end{table}

\begin{figure}[H]
\centering
\includegraphics[width=0.55\textwidth]{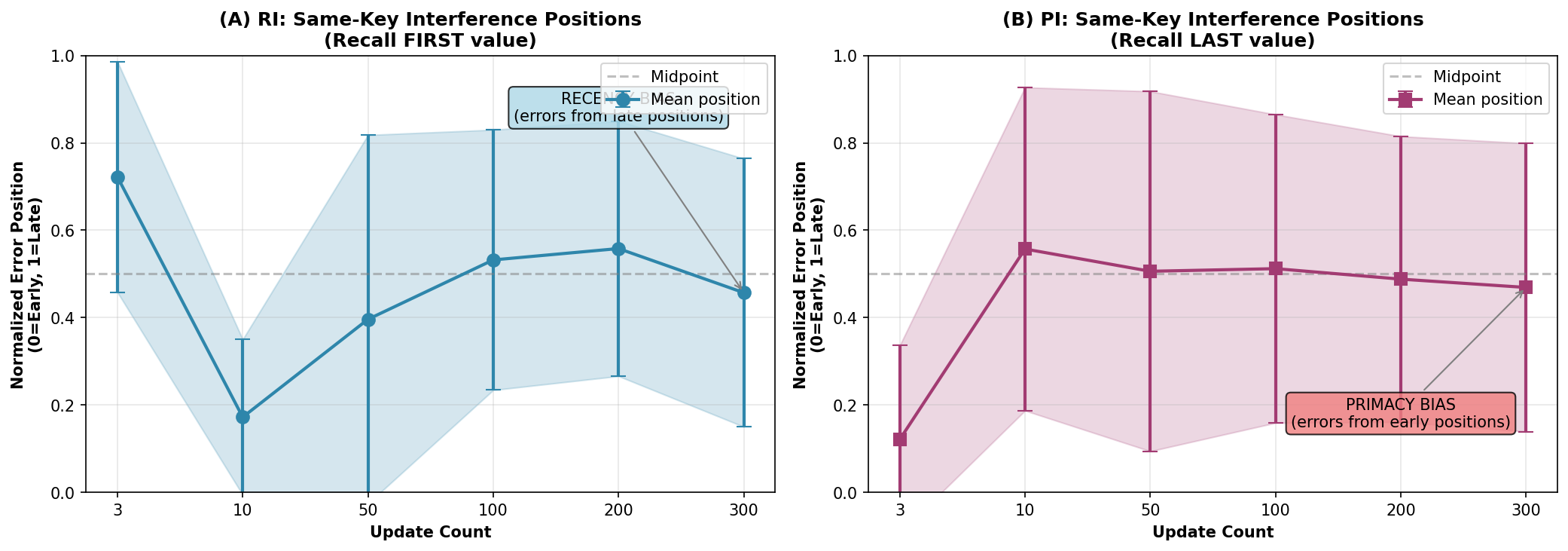}
\vspace{-4pt}
\caption{\textbf{Positional Origins of Intrusion Errors.} At $N$=3, RI intrusions originate from late positions (0.72); PI from early positions (0.12). Both converge toward 0.5 as $N$ grows.}
\label{fig:position_bias}
\end{figure}

\end{document}